# Calibrating DFT formation enthalpy calculations by multi-fidelity machine learning

Sheng Gong[1], Shuo Wang[2], Tian Xie[3], Woo Hyun Chae[1], Runze Liu[1], and Jeffrey C. Grossman[1,*]

[1]Department of Materials Science and Engineering, Massachusetts Institute of Technology, MA 02139, USA

[2]Department of Materials Science and Engineering, University of Maryland, MD 20742, USA

[3]Computer Science and Artificial Intelligence Lab, Massachusetts Institute of Technology, MA 02139, USA

# ABSTRACT

Machine learning materials properties measured by experiments is valuable yet difficult due to the limited amount of experimental data. In this work, we use a multi-fidelity random forest model to learn the experimental formation enthalpy of materials with prediction accuracy higher than the empirically corrected PBE functional and meta-GGA functionals (PBEsol, SCAN and r$^2$SCAN), and it outperforms the hotly studied deep neural-network based representation learning and transfer learning. We then use the model to calibrate the DFT formation enthalpy in the Materials Project database, and discover materials with underestimated stability. The multi-fidelity model is also used as a data-mining approach to find how DFT deviates from experiments by the explaining the model output.

# INTRODUCTION

In order to accelerate the design of new materials, accurate computational methods such as Density Functional Theory (DFT)[1] have been employed to generate large datasets that contain more than $10^5$ entries of materials properties, including the Materials Project (MP)[2], Open Quantum Materials Database (OQMD)[3], the Automatic Flow of Materials Discovery Library (AFLOW)[4], and the Joint Automated Repository for Various Integrated Simulations (JARVIS)[5]. While the availability of such databases has boosted the exploration of novel materials[6, 7, 8, 9, 10, 11, 12, 13, 14, 15], it is important to note that most of the data is generated with computationally "cheap" DFT functionals such as PBE[16], that can in turn lead to non-negligible errors when compared with experimental measurements.

As an example, the formation enthalpy ($\Delta H_f$) is a fundamental property that determines the thermodynamic stability of materials. The mean absolute error (MAE) between the computed $\Delta H_f$ in these large DFT databases and experimental measurements are reported to be ~0.1 eV/atom[3, 17]. Due to the sensitivity of phase stability to energy, such a difference (~0.1 eV/atom) might be the difference between a material that is readily synthesizable and one that is almost impossible to realize[18, 19, 20]. In addition, because of the limited amount of available experimental data, currently most machine learning (ML) models applied to materials are trained on DFT datasets[6, 21, 22, 23, 24, 25, 26, 27, 28, 29, 30, 31, 32, 33, 34, 35], making any error in the DFT calculations critical to the usefulness of such ML models[7, 31, 36, 37].

To improve the accuracy of formation enthalpy calculations, a number of density functionals have been developed, such as PBEsol[38], SCAN[39], r$^2$SCAN[40] and HSE[41], which have shown significant improvement in accuracy of formation enthalpy calculation[42, 43, 44]. On the other hand, these more accurate functionals are also computationally more expensive, limiting their utility for generation of large databases[43, 45]. Empirical corrections represent another, faster approach to improve the accuracy of prediction of $\Delta H_f$. For example, in the MP dataset, $\Delta H_f$ of certain materials (including oxides, phosphates, borates and silicates) is empirically corrected by fitted element corrections[46], and in OQMD $\Delta H_f$ is corrected by a chemical-potential fitting[3]. Very recently, Wang *et al.*[47] proposed a linear correction scheme with error of 0.051 eV/atom compared with experimental values on a dataset with 222 materials containing certain anions and transition metals. Yet, despite this recent success in lowering the error for some chemical systems[48], such corrections are based on human understanding of specific chemistries and relatively simple assumptions, and are thus difficult to be transferrable across different chemistries[46, 48]. It would be beneficial to design prediction schemes that can automatically extract chemistry-property relationship across different chemistries without human intervention, and data-driven ML methods[18, 24, 26, 28, 29, 32, 45] are promising candidates to learn the complex mapping between chemistry and $\Delta H_f$.

One of the biggest challenges in machine learning materials properties is the lack of experimental data[49]. Efforts have been made to improve the performance of learning on small experimental datasets by extracting and transferring information from large DFT datasets. Currently, there are mainly two strategies to achieve the transfer between DFT and experimental datasets, transfer learning[21, 28, 50, 51, 52, 53] and multi-fidelity machine learning[45, 54, 55, 56]. The idea of transfer learning (see Figure 1a) is first learning large DFT datasets (source) using a large neural network, and then transferring the weights of the network to the machine learning task of small experimental datasets (target). Although transfer learning has achieved success in problems where the source and target datasets are highly correlated[21, 28, 50, 51], the approach is mostly applied to neural network architectures, and if the correlation is not strong enough, transfer learning will not improve and may even deteriorate the learning performance[52]. Different from transfer learning where information is passed by transferring network parameters, in multi-fidelity machine learning (see Figure 1b) information of cheap and low-fidelity data is directly passed to the learning task of expensive and high-fidelity data, either in the feature (input) level[54] or in the label (output) level[45, 55, 56, 57]. In other words, the low-fidelity data can be used as feature in the machine learning task of high-fidelity data, or the task of machine learning the high-fidelity data can be converted to the task of machine learning the difference between high-fidelity data and low-fidelity data, which is also known as Δ-Machine Learning[57]. From the handful of previous studies, multi-fidelity machine learning has shown higher predictive power than the single-fidelity ones (directly learning the high-fidelity data) on materials properties like band gaps and energies from different density functionals[45, 55, 56, 57]. However, there is no previous work that adapt multi-fidelity machine learning in both feature and label level at the same time.

In this work, we present a comprehensive machine learning study about $\Delta H_f^{exp}$ using transfer learning and multi-fidelity machine learning. For the machine learning architectures, we compare four different models, random forests (RF), multi-layer perceptron (MLP), Representation Learning from Stoichiometry (ROOST)[26] and Crystal Graph Convolutional Neural Network (CGCNN)[32]. We find that multi-fidelity RF in both the feature and label level has the best prediction performance for $\Delta H_f^{exp}$ with a half reduction in MAE compared with DFT results from MP, and improved performance compared to recent linear correction schemes[47] as well as more sophisticated density functionals like PBEsol[38], SCAN[39] and r$^2$SCAN[40]. We also analyze the effects of machine learning architectures, featurization methods and information transfer strategy on learning $\Delta H_f^{exp}$ and $\Delta H_f^{diff}$. Further, the more accurate $\Delta H_f$ are applied to re-evaluate the thermodynamic stability of materials, and cases with underestimated stability in the MP database are discovered. We also use the machine learning model to find where current DFT deviates from experiments by explaining the model output.

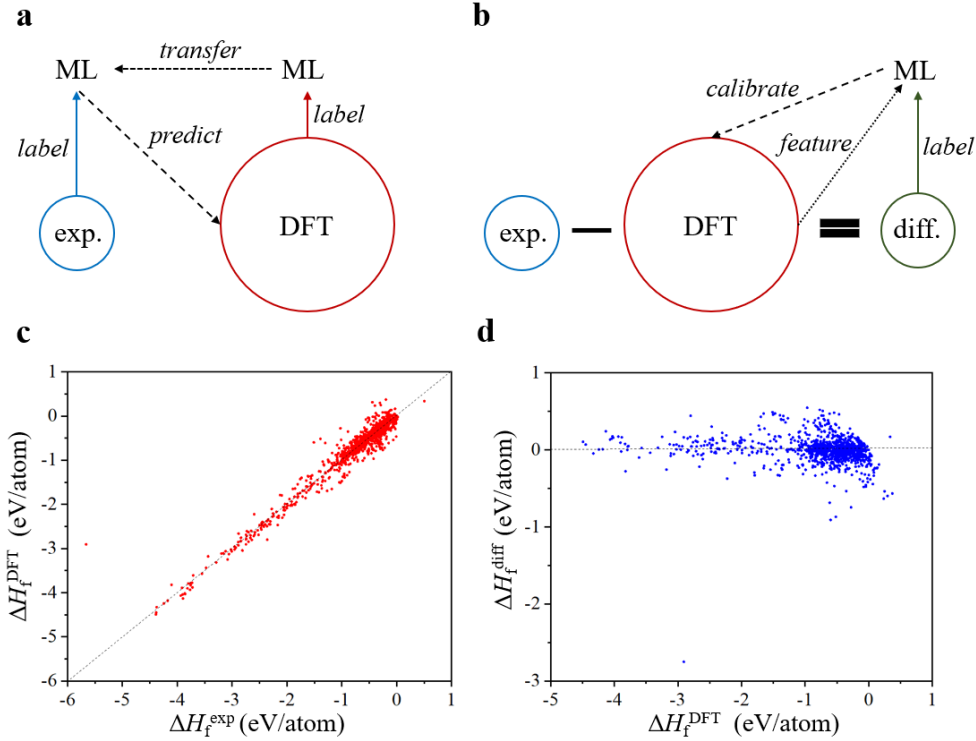

**Figure 1. Illustrations of the machine learning frameworks and datasets used in this work. a** and **b** Schematics of transfer learning and multi-fidelity machine learning in this work, respectively. In **a**, the $\Delta H_f^{DFT}$ are first used as label to train a ML model, then the weights of the first ML model are transferred to initialize a second ML model, and the $\Delta H_f^{exp}$ are used as label to train the second model, finally the second model is used to predict $\Delta H_f^{exp}$ of all materials in the large DFT dataset. In **b**, first the dataset of the difference between $\Delta H_f^{exp}$ and $\Delta H_f^{DFT}$ are constructed ($\Delta H_f^{diff}$), then $\Delta H_f^{diff}$ are used as label to train a ML model with the $\Delta H_f^{DFT}$ as an input feature, and finally the trained model is used to calibrate the different between $\Delta H_f^{DFT}$ and $\Delta H_f^{exp}$ for all materials in the large DFT dataset. **c** $\Delta H_f^{DFT}$ versus $\Delta H_f^{exp}$. **d** $\Delta H_f^{diff}$ versus $\Delta H_f^{DFT}$.

## RESULTS

**Illustration of machine learning frameworks and datasets.** In this work, we use two different strategies to learn $\Delta H_f^{exp}$ with the assistance of information from the MP dataset, transfer learning and multi-fidelity machine learning (in the following, "$\Delta H_f^{DFT}$" denotes the empirically-corrected PBE $\Delta H_f$ by Jain *et al.*[46] from the MP database, V2021.03.22). As shown in Figure 1a, in transfer learning a neural network is first trained on the large MP dataset with more than $10^5$ data points of $\Delta H_f^{DFT}$, then weights of the neural network are transferred to initialize a second neural network, and finally part of the weights of the second network are optimized by the small $\Delta H_f^{exp}$ dataset. Once trained, the second neural network can serve to predict $\Delta H_f^{exp}$ of materials in the large MP dataset. In multi-fidelity machine learning, as shown in Figure 1b, first the dataset of $\Delta H_f^{diff}$ ($\Delta H_f^{exp}$ - $\Delta H_f^{DFT}$) is built, then machine learning models are trained on $\Delta H_f^{diff}$ dataset, and in the training process, $\Delta H_f^{DFT}$ can also serve as an input feature of each material. Once trained, the machine learning model can serve to calibrate the $\Delta H_f^{DFT}$ by adding $\Delta H_f^{diff}$ to $\Delta H_f^{DFT}$ to get the $\Delta H_f^{exp}$. The key difference between transfer learning and multi-fidelity machine learning is that in the former two networks are trained and information transfer is achieved by transferring network weights, while in the later only one model is trained and information transfer is achieved by learning the difference between two datasets and adding the $\Delta H_f^{DFT}$ as one of the input features. In addition to the two basic strategies as shown in Figure 1a and 1b, variants are also tested in this work, including combination of transfer learning and multi-fidelity machine learning (initializing a network from one trained on $\Delta H_f^{DFT}$ and optimizing the newly initialized network by $\Delta H_f^{diff}$), and multi-fidelity machine learning by only learning $\Delta H_f^{diff}$ or only adding $\Delta H_f^{DFT}$ as input feature.

As described above, we choose four different machine learning architectures to realize transfer learning and/or multi-fidelity machine learning, which are RF, MLP, ROOST and CGCNN. The choice aims to increase the variety of machine learning architectures to fairly evaluate the effect of transfer learning and multi-fidelity learning, and to enlarge the hypothesis space to search for the best machine learning models for predicting $\Delta H_f^{exp}$. These ML architectures also provide varieties in terms of basic algorithms, input information and featurization: MLP, ROOST and CGCNN are based on neural networks while RF is not; ROOST only needs compositions as input while CGCNN takes both compositions and 3D structures as input, and RF and MLP can be trained either with or without structural information; RF and MLP need human-engineered featurization while ROOST and CGCNN learn fingerprints of materials in the training process.

In this work, we choose the Materials Project database (MP, V2021.03.22) as the source of $\Delta H_f^{DFT}$, because MP is a widely used large DFT database, and the difference of $\Delta H_f$ between MP and other large DFT databases is not large. For example, the difference between $\Delta H_f$ of 563 materials from MP and OQMD is reported to be 0.028 eV/atom[3]. As for the experimentally measured $\Delta H_f$, we combine the IIT dataset[17] and SSUB dataset[58] and remove the duplicates, leading to 1143 data points with available $\Delta H_f^{exp}$, $\Delta H_f^{DFT}$, and DFT optimized 3D atomic structures from MP. In addition to the value of $\Delta H_f^{exp}$, there are also uncertainty estimations in the IIT dataset[17], from which one can see that the mean uncertainty of $\Delta H_f^{exp}$ based on 499 materials is around 0.023 eV/atom. More details about the data collection procedure are provided in the **METHODS** section. $\Delta H_f^{DFT}$ and $\Delta H_f^{exp}$ are compared in Figure 1c, from which one can see that $\Delta H_f^{DFT}$ are already quite close to $\Delta H_f^{exp}$ in value, and there is no clear systematic shift between $\Delta H_f^{DFT}$ and $\Delta H_f^{exp}$. As shown in Figure 1d, the distribution of $\Delta H_f^{diff}$ is centered around zero, and there is no obvious correlation between $\Delta H_f^{diff}$ and $\Delta H_f^{DFT}$. From Figure 1c and 1d, one can see that $\Delta H_f^{diff}$ has a narrower distribution than $\Delta H_f^{exp}$ with the standard deviation of 0.1718 eV/atom and 0.8000 eV/atom for the $\Delta H_f^{diff}$ dataset and $\Delta H_f^{exp}$ dataset, respectively.

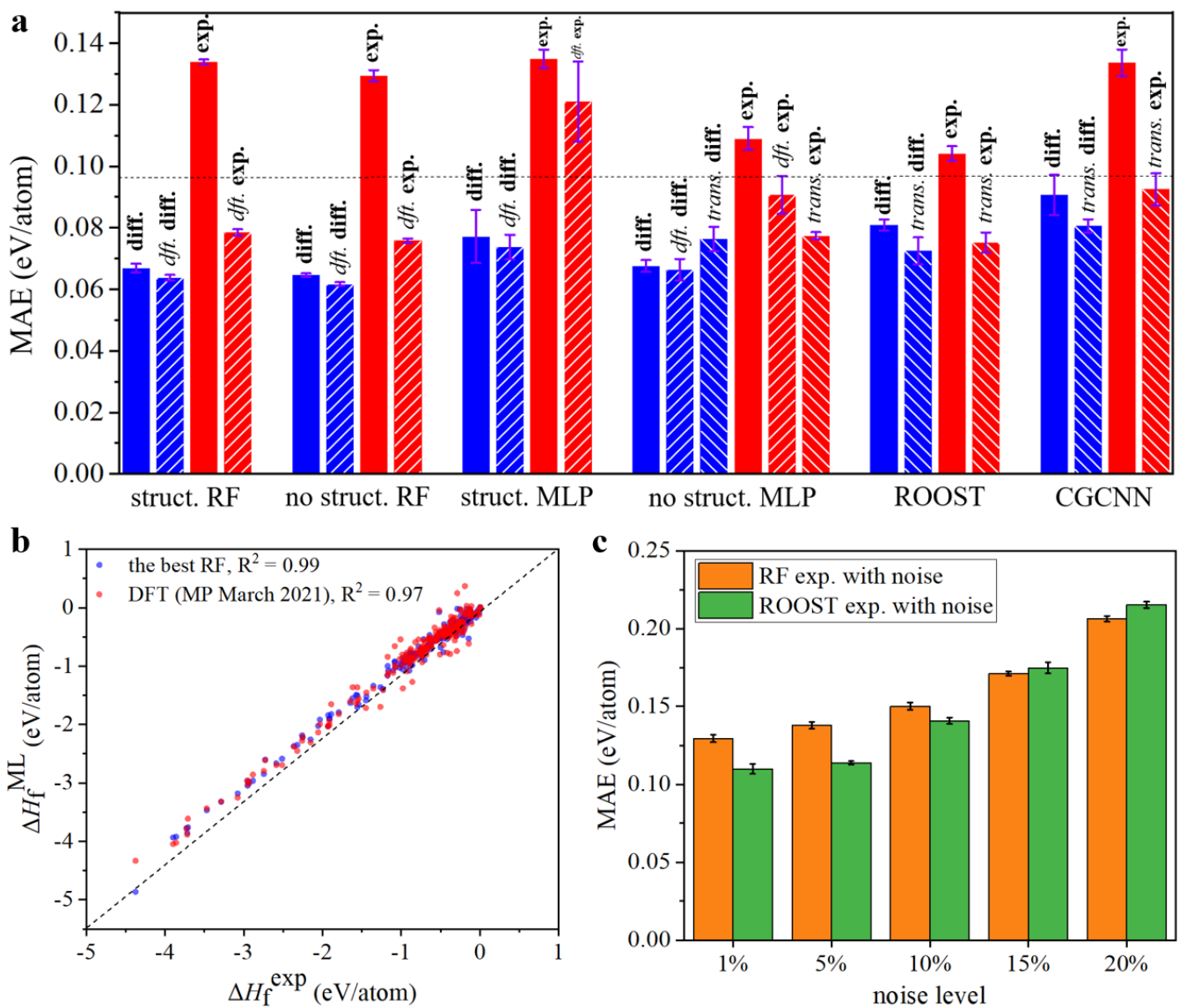

**Figure 2. Comparison of machine learning models. a** Mean average errors (MAE) between predictions of $\Delta H_f$ from machine learning models and experimental measurements. Each type of machine learning model is trained 10 times to estimate the uncertainty levels. RF denotes random forest, MLP denotes multilayer perceptron, and ROOST[26] and CGCNN[32] are two deep-learning models that automatically extract materials' fingerprints from compositions and structures, respectively. Here, "struct." means the model is trained with structural and compositional features, "no struct." the model is trained with only compositional features, "dft." the model is trained with $\Delta H_f^{DFT}$ as an input, "trans." the model is trained

in a transfer learning manner, "diff." the model is trained on $\Delta H_f^{diff}$, "exp." the model is directly trained $\Delta H_f^{exp}$. The dashed horizontal line corresponds to the MAE of $\Delta H_f^{DFT}$. **b** $\Delta H_f^{exp}$ versus $\Delta H_f^{ML}$ from the best RF model (the sixth from the left in **a**) and $\Delta H_f^{DFT}$. **c** MAE of predictions $\Delta H_f^{exp}$ with noise from RF and ROOST. Under each noise level, gaussian noises with standard deviation of noise level*0.8 eV/atom (0.8 eV/atom is the standard deviation of the $\Delta H_f^{exp}$ dataset) are added to both training set and test set.

**Predicting $\Delta H_f^{exp}$ by machine learning.** For the RF and MLP, compositional and structural features are provided from matminer[59] as input features (a list of features is provided in the **METHODS** section), for ROOST only the compositions are provided as input and it automatically learns the fingerprints of materials, and for CGCNN the compositions and 3D atomic structures are provided as input and the fingerprints are learned in the training. In order to test the prediction performance, 20% of the 1143 materials are randomly chosen as the test set. Details about the training procedure are provided in the **METHODS** section. As a baseline, for the test set, we find that the MAE between $\Delta H_f^{DFT}$ and $\Delta H_f^{exp}$ is 0.0955 eV/atom. The test results for all machine learning models are shown in Figure 2a with a detailed list in Table S1, and here we analyze the results from the following aspects:

(1). The best performance is achieved with the RF model that is trained on $\Delta H_f^{diff}$ and has compositional features and $\Delta H_f^{DFT}$ as input features (Figure 2a). The error for this best case is roughly 30% lower than that of $\Delta H_f^{DFT}$, with a level of error (0.0617 eV/atom). The parity plot of $\Delta H_f^{DFT}$ and $\Delta H_f^{ML}$ from the best RF model versus $\Delta H_f^{exp}$ of the test set is shown in Figure 2b, from which one can observe that $\Delta H_f$ from the best RF model aligns closer to the $\Delta H_f^{exp}$ than $\Delta H_f^{DFT}$ within the range from -5 eV/atom to 1 eV/atom. Predictions from the best RF model also have a higher $R^2$ score (0.99) than that from the DFT calculations in the MP database (0.97).

Very recently, Kingsbury *et al.*[44] performed high-throughput calculations for 6,000 materials by PBEsol[38], SCAN[39] and r$^2$SCAN functional[40], which are three meta-GGA functionals that are expected to have higher accuracy and higher computational cost than GGA functional such as PBE[43]. In Table 1, MAEs between experimental $\Delta H_f$ and $\Delta H_f$ from different density functionals with different empirical corrections are listed. Note that, different from Figure 2, the reported MAEs in Table 1 are based on a dataset with 122 materials that have all the values of $\Delta H_f$ from different sources (these materials are in the test set mentioned above). One can observe that, the best RF model in this work outperforms the three meta-GGA functionals and the two empirical corrections. MAE of the best RF model is almost half of that of SCAN[39], PBEsol[38], and also almost half of that of the corrections from Jain *et al.*[46] and Wang *et al.*[47]. The superiority of the best RF model over the meta-GGA functionals is encouraging, because i) the best RF model provides lower error compared with more sophisticated density functionals, ii) it is much faster than the self-consistent DFT simulations, especially with meta-GGA functionals, enabling one to screen $\Delta H_f$ of materials accurately in a high-throughput fashion. For example, for the $10^5$ materials in large DFT databases such as MP, more accurate predictions of $\Delta H_f$ can be calculated by the RF models within minutes, while that from meta-GGA functionals may take months of calculations. Note that for new materials without low-fidelity $\Delta H_f$ predictions yet (such as corrected-PBE), computational cost for the low-fidelity $\Delta H_f$ should be added to the total cost of the best RF model. For scenarios where evaluations of $\Delta H_f$ for new materials with cost lower than PBE are required, as shown in Figure 2a, "no. struct. MLP" with transfer learning and ROOST[26] with transfer learning are recommended machine learning models.

As for the superiority of the best RF model over the recent linear correction scheme from Wang *et al.*[47] as shown in Table 1, there are four possible explanations: i) the RF model takes non-linear effects into account, ii) the compositional descriptors used here capture more information than simple stoichiometry used in Wang *et al.*[47], iii) the learned correction in Wang *et al.*[47] is only from materials with certain anions and transition metals while in the present work there is no such constraint, and iv) the calibration scheme used here is built on empirically corrected PBE results as opposed to uncorrected PBE data in Wang et al.[47].

Table 1. Comparison of MAEs between $\Delta H_f^{exp}$ and $\Delta H_f$ from different density functionals with different corrections. Different from Figure 2, the reported MAEs here are based on a dataset with 122 materials in the test set that have all the values of $\Delta H_f$ from different sources. The two corrections in the cell of "PBE (Jain et al.[46], the best RF)" show that the PBE $\Delta H_f$ is first corrected by Jain et al.[46] then corrected by the best RF model in this work. "(no)" in the right three cells at the upper row means that no correction is applied to the $\Delta H_f$ from the density functional. "PBE (Jain et al.[46])" is the one used in the MP database before May 2021 (V2021.03.22) and is the one used as the low fidelity data in this work ("$\Delta H_f^{DFT}$"). "PBE (Wang et al.[47])" is the one used in the MP database after May 2021 (V2021.05.13). MAE is in the unit of eV/atom.

| Functional (Correction) | PBE (Jain et al.[46], the best RF) | PBE (Jain et al.[46]) | PBE (Wang et al.[47]) | PBEsol (no)[44] | SCAN (no)[44] | r$^2$SCAN (no)[44] |
|---|---|---|---|---|---|---|
| MAE | 0.0542 | 0.0935 | 0.0927 | 0.0973 | 0.1024 | 0.0825 |

(2). Training the machine learning models on $\Delta H_f^{diff}$ helps to reduce error compared with training models on $\Delta H_f^{exp}$ directly, as under the same condition (architecture and featurization), the models trained on $\Delta H_f^{diff}$ always have lower MAE than that trained on $\Delta H_f^{exp}$. Here, we attribute the lower absolute error of learning $\Delta H_f^{diff}$ to the fact that $\Delta H_f^{diff}$ has a narrower distribution than $\Delta H_f^{exp}$ with 5 times smaller standard deviation (0.17 eV/atom versus 0.80 eV/atom). One can imagine that, if $\Delta H_f^{diff}$ and $\Delta H_f^{exp}$ have the same distribution except a scaling factor of 1/5, then ideally the MAEs of ML models (with proper normalization) trained on $\Delta H_f^{diff}$ should also be 1/5 of that trained on $\Delta H_f^{exp}$. However, the MAEs of models trained on $\Delta H_f^{diff}$ are all larger than 1/5 of that trained on $\Delta H_f^{exp}$, suggesting that $\Delta H_f^{diff}$ is easier to learn absolutely but harder to learn relatively than $\Delta H_f^{exp}$.

In order to further illustrate the above explanation, we use $R^2$ score, a unitless metric invariant to scaling, to show the relative difficulty of predicting $\Delta H_f^{diff}$ and $\Delta H_f^{exp}$. The $R^2$ of predictions of $\Delta H_f^{diff}$ by the best RF model is 0.54 (here $R^2$ of 0.54 is based on predicted $\Delta H_f^{diff}$ versus true $\Delta H_f^{diff}$, while the $R^2$ of 0.99 in Figure 2b is based on predicted $\Delta H_f^{exp}$ versus true $\Delta H_f^{exp}$), while the $R^2$ of predictions of $\Delta H_f^{exp}$ by the same RF model is 0.94, suggesting that $\Delta H_f^{exp}$ is easier to learn relatively than $\Delta H_f^{diff}$.

(3). Feeding $\Delta H_f^{DFT}$ as one of the input features helps to lower the error. As with the same machine learning architecture (RF or MLP), label, and other features, models with $\Delta H_f^{DFT}$ as one of the input features always have lower error than that without $\Delta H_f^{DFT}$. This effect is more significant when the models are trained on $\Delta H_f^{exp}$, because as shown in Figure 1c $\Delta H_f^{DFT}$ has a strong correlation with $\Delta H_f^{exp}$, while as shown in Figure 1d the correlation between $\Delta H_f^{DFT}$ and $\Delta H_f^{diff}$ is not obvious.

Combining analysis (2) and (3), one can observe that, adapting the strategy of multi-fidelity machine learning might help to significantly lower prediction error, if the difference between the different fidelity datasets has a narrower distribution than the high-fidelity dataset, and/or if there is a strong correlation between the different fidelity datasets. Machine learning models with both the modifications of changing label and adding extra input features might outperform that with either single modification.

(4). Similar to (3), transfer learning helps more when transferring from $\Delta H_f^{DFT}$ to $\Delta H_f^{exp}$ than from $\Delta H_f^{DFT}$ to $\Delta H_f^{diff}$ because of the stronger correlation between $\Delta H_f^{DFT}$ and $\Delta H_f^{exp}$. On the other hand, transfer learning helps more for the deep learning architectures that learn the representations of materials (ROOST and CGCNN), confirming that the power of transfer learning in materials science lies on the transfer of learned materials representations from larger datasets[21, 53]. On the other hand, it helps less or even hurts performance of MLP with *off-the-shelf* featurization, suggesting that transferring the mapping function between representations and properties might lead to negative transfer[60] if the correlation between source and target is not very strong, such as the case of "no struct. MLP" in Figure 2a.

(5) RF with human-engineered features performs better than ROOST and CGCNN, two deep representation learning models, when trained on $\Delta H_f^{diff}$, while RF performs similar or worse than neural-network based models when trained on $\Delta H_f^{exp}$. Although it is not surprising that neural-network based deep learning algorithms don't show superior performance over RF due to the limited dataset size[52, 61], the effect of learning targets ($\Delta H_f^{diff}$ and $\Delta H_f^{exp}$) on prediction performance of different machine learning models is interesting and worth of being discussed.

The different uncertainty level between $\Delta H_f^{diff}$ and $\Delta H_f^{exp}$ might help to explain why RF performs better than neural network-based models when trained on $\Delta H_f^{diff}$ while there is no such superiority of RF when trained on $\Delta H_f^{exp}$. As discussed above, $\Delta H_f^{diff}$ has a narrower distribution than $\Delta H_f^{exp}$. Because $\Delta H_f^{diff} = \Delta H_f^{exp} - \Delta H_f^{DFT}$, if we consider $\Delta H_f^{exp}$ and $\Delta H_f^{DFT}$ as two independent random variables, then $\Delta H_f^{diff}$ should have larger uncertainty than $\Delta H_f^{exp}$. Therefore, the robustness of RF against uncertainty[61, 62] might explain the superiority of RF when trained on $\Delta H_f^{diff}$. The larger uncertainty level of $\Delta H_f^{diff}$ might also help to explain why $\Delta H_f^{diff}$ is harder to learn relatively than $\Delta H_f^{exp}$ as in (2).

In order to further investigate the effect of uncertainty on performance of machine learning models, RF and ROOST are employed to learn $\Delta H_f^{exp}$ with random noises, a source of uncertainty. In Figure 2a, one can see that RF performs worse than ROOST when trained on $\Delta H_f^{exp}$. In Figure 2c, the MAEs of RF and ROOST and the corresponding noise levels are shown. One can see that, under low noise levels the errors of RF are still higher than that of ROOST, while under high noise levels the errors of RF become lower than that of ROOST. The different relative performance of RF and ROOST under different noise levels agrees with the superiority of RF against uncertainty[61, 62], and supports our hypothesis that the different uncertainty levels of the $\Delta H_f^{diff}$ dataset and the $\Delta H_f^{exp}$ dataset might explain why RF is better on the $\Delta H_f^{diff}$ dataset while ROOST is better on the $\Delta H_f^{exp}$ dataset.

Based on the fact that when trained on $\Delta H_f^{diff}$, random forest with human-engineered featurization outperforms neural networks-based models, especially deep representation learning models, we suggest that for machine learning applications in the field of materials science, with limited dataset size and without proof of a low uncertainty level of the dataset, deep neural network-based representation learning algorithms[24, 26, 32, 63] should not be the only type of models employed, and other feature engineering methods and machine learning architectures beyond neural networks should also be tested.

While there are some previous works show that information of local bonding environment can be used to correct formation enthalpies of certain materials like sulfides[64], fluorides[65] and oxides[48, 65], in this work, the machine learning models with only compositions as input outperform those with both compositions and structures as input. One of the possible causes of the phenomenon is that there still lacks the data points of polymorphs with the same composition but different $\Delta H_f^{exp}$ in the current dataset, which suggests the urgency of building a comprehensive $\Delta H_f^{exp}$ dataset with sufficient entries of polymorphs to comprehensively understand the role of structures in determining $\Delta H_f^{exp}$.

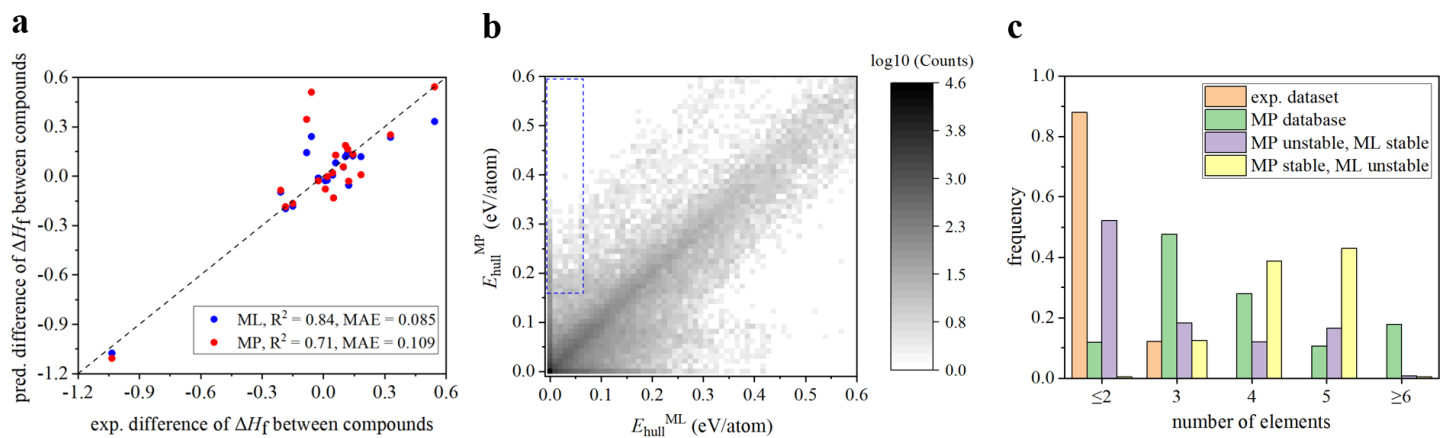

**Figure 3. Stability evaluation from energy above hull. a** Difference of $\Delta H_f$ between pairs of compounds in the same chemical system from experiments versus that from MP and machine learning. **b** Distribution of energy above hull ($E_{hull}$, in eV/atom) of all materials in the Materials Project[2] database calculated by the corrected-PBE $\Delta H_f$ in MP ($E_{hull}^{MP}$) versus that calculated by the machine learning $\Delta H_f$ in this work ($E_{hull}^{ML}$). Here, $E_{hull}$ is constructed from all materials in the Materials Project database. The color scheme is used to show the (log10 of) number of materials within a range of certain $E_{hull}^{ML}$ and $E_{hull}^{MP}$, and the red rectangle shows the area with $E_{hull}^{MP} > 0.16$ eV/atom and $E_{hull}^{ML} < 0.06$ eV/atom. **c** Appearance frequencies of number of elements of each material in the datasets. Here, "exp. dataset" is the $\Delta H_f^{exp}$ used in this work, "MP database" is the set of all materials in the Materials Project database, "MP unstable, ML stable" is the set of materials with $E_{hull}^{MP} > 0.16$ eV/atom and $E_{hull}^{ML} < 0.06$ eV/atom and "MP stable, ML unstable" is the set of materials with $E_{hull}^{MP} < 0.06$ eV/atom and $E_{hull}^{ML} > 0.16$ eV/atom.

**Discovering materials with underestimated stability in MP.** With the best RF model that can significantly lower the error of $\Delta H_f$ from the MP database, we can calibrate $\Delta H_f$ of all materials in the MP database. The dataset with all the calibrated $\Delta H_f$ is provided in the **DATA AVAILABILITY** section, and as an application, here we use the calibrated $\Delta H_f$ to re-evaluate the thermodynamic stability of all materials in the MP database by constructing the energy above hull ($E_{hull}$, the energy difference between the candidate compound and the ground-state phase(s) in a compositional space[66]. More discussions about $E_{\text{hull}}$ are provided in the **METHODS** section and Figure S3). However, as Bartel et al.[18] pointed out, although sometimes DFT has large errors for prediction of $\Delta H_f$, $\Delta H_f^{\text{DFT}}$ of similar materials contain similar systematic errors, and when evaluating phase stability, the cancellation of systematic errors makes DFT more useful for evaluating relative stability between compounds than some machine learning models with similar or even better accuracy with respect to $\Delta H_f^{\text{exp}}$.

Therefore, before screening $E_{\text{hull}}$ for the full MP dataset, we first evaluate the performance of $\Delta H_f^{\text{DFT}}$ and $\Delta H_f^{\text{ML}}$ for evaluating relative stability between compounds. Since there are only 229 materials in the test set, which are not enough for constructing phase diagrams and $E_{\text{hull}}$, we use the difference between $\Delta H_f$ of pairs of compounds in the same chemical system to evaluate relative stability between compounds. We list all 20 pairs of compounds in the same chemical system in the test set in Table 2, and we also plot the difference from experiments versus that from MP and machine learning (ML) in Figure 3a. One can see that ML outperforms MP in terms of difference of $\Delta H_f$ between compounds in the same chemical system, which shows that the ML model outperforms DFT for relative stability evaluation.

Table 2. Difference of $\Delta H_f$ between pairs of compounds in the same chemical system from different sources. Difference of $\Delta H_f$ is the unit of eV/atom.

| Pair of Compounds | Experiment | Materials Project | Machine Learning in This Work |
|---|---|---|---|
| $TiFe_2$ - $TiFe$ | 0.0487 | -0.1324 | -0.1316 |
| $BiI_3$ - $BiI$ | 0.1075 | 0.1868 | 0.1193 |
| $LuIr_2$ - $LuIr$ | -0.1502 | -0.1664 | -0.1826 |
| LaSi - La$_5$Si$_3$ | 0.143 | 0.1335 | 0.1229 |
| BMo$_2$ - BMo | -0.1858 | -0.1856 | -0.1972 |
| Na$_2$O - NaO$_2$ | 0.5435 | 0.5428 | 0.3328 |
| BW$_2$ - B$_5$W$_2$ | -0.0591 | 0.5108 | 0.2408 |
| Co$_3$O$_4$ - CoO | 0.1229 | -0.0302 | -0.0553 |
| ZrCo$_2$ - Zr$_2$Co | 0.0974 | 0.0574 | 0.0553 |
| TmAg - TmAg$_2$ | 0.1835 | 0.0088 | 0.1187 |
| PrNi$_5$ - PrNi | -0.0259 | -0.0281 | -0.0116 |
| TiAu$_2$ - TiAu | 0.0179 | -0.0026 | -0.0243 |
| NdRh - NdRh$_2$ | 0.0446 | 0.0202 | 0.0064 |
| CaO$_2$ - CaO | -1.0353 | -1.1070 | -1.075 |
| Zr$_5$Si$_3$ - Zr$_5$Si$_4$ | -0.2094 | -0.0855 | -0.0964 |
| Zr$_5$Si$_3$ - ZrSi$_2$ | 0.1181 | 0.1654 | 0.1397 |
| Zr$_5$Si$_4$ - ZrSi$_2$ | 0.3275 | 0.2509 | 0.2361 |
| Mn$_2$Sb - MnSb | -0.0824 | 0.3453 | 0.1428 |
| CrSi - CrSi$_2$ | 0.0090 | -0.0783 | -0.0280 |
| Mn$_{11}$Si$_{19}$ - Mn$_3$Si | 0.0596 | 0.1276 | 0.0809 |

We next re-evaluate materials stability using ML calibrated $\Delta H_f$ to construct $E_{hull}^{ML}$ for all materials in the MP database using all compositions in MP. In chemical intuition, materials with smaller $E_{hull}$ tend to be more thermodynamically synthesizable and stable[19, 20, 67], although $E_{hull} = 0$ is not a hard threshold for successful synthesis and room-temperature and pressure stability of materials because of other factors such as kinetics[68], and in practice empirical heuristics of several room temperature $k_B$T are used as stability thresholds[19, 20, 67]. In Figure 3b the distributions of $E_{\text{hull}}$ of all materials in the MP database constructed from $\Delta H_f^{\text{DFT}}$ and $\Delta H_f^{\text{ML}}$ of all compositions in the MP database are shown, from which one can see that most materials have similar $E_{\text{hull}}^{\text{MP}}$ and $E_{\text{hull}}^{\text{ML}}$, and majority of materials have close-to-zero $E_{\text{hull}}^{\text{MP}}$ and $E_{\text{hull}}^{\text{ML}}$. More importantly, there are materials with large $E_{\text{hull}}^{\text{MP}}$ and small $E_{\text{hull}}^{\text{ML}}$. These materials might have underestimated stabilities in MP. For example, there are 800 materials in the red rectangles in the upper-left corner in Figure 3b that have $E_{\text{hull}}^{\text{MP}} > 0.16$ eV/atom and $E_{\text{hull}}^{\text{ML}} < 0.06$ eV/atom, among which there are around 100 already synthesized materials. (The thresholds are set to be relaxed from 6 times and 2 times of room Temperature $k_B T$[19, 20]). As examples, we list some interesting materials in Table 3 with novel physical properties and/or potential applications with $E_{\text{hull}}^{\text{MP}} > 0.16$ eV/atom and $E_{\text{hull}}^{\text{ML}} < 0.06$ eV/atom, where there are both synthesized materials and hypothetical materials. One can see that there are a number of materials with various applications ranging from battery electrodes[69], catalysts[70, 71, 72] to optical[73, 74, 75], electronic[76, 77], magnetic[78, 79, 80, 81, 82] devices and superconductors[83, 84], for which $E_{\text{hull}}^{\text{ML}}$ succeeds in explaining their synthesizability while $E_{\text{hull}}^{\text{MP}}$ does not. One extreme example is MnSnIr[85], a stable Half-Heusler compound synthesized from a peritectic reaction[86], of which $E_{\text{hull}}^{\text{MP}}$ is considerably high (0.5117 eV/atom) while $E_{\text{hull}}^{\text{ML}}$ is 0, showing the utility of our machine learning model to identify and correct underestimated stabilities in the MP database. In addition to the already synthesized materials, those unrealized hypothetical materials provide potential opportunities for energy and environmental materials[87, 88, 89], structural materials[90] and electronic devices[91, 92], and as shown in Table 3 and Figure 3b, many of these materials that are estimated stable by $E_{\text{hull}}^{\text{ML}}$ might have underestimated stability in the MP database. Therefore, in the future, if experimentalists intend to realize those materials, large $E_{\text{hull}}^{\text{MP}}$ alone should not be sufficient for excluding the trial of synthesis if those materials have small $E_{\text{hull}}^{\text{ML}}$.

Note that there are also 1,000 materials in the lower-right corner in Figure 3b that have $E_{\text{hull}}^{\text{MP}} < 0.06$ eV/atom and $E_{\text{hull}}^{\text{ML}} > 0.16$ eV/atom. Details of those materials can be obtained in the shared online dataset.

An extreme example is LiNbGeO$_5$,[93] a synthesized compound with $E_{\text{hull}}^{\text{MP}}$ of 0 and $E_{\text{hull}}^{\text{ML}}$ of 0.4334 eV/atom. The $\Delta H_{\text{f}}^{\text{ML}}$ and $E_{\text{hull}}^{\text{ML}}$ of all materials in the MP database are provided in the **DATA AVAILABILITY** section.

In order to further investigate how MP and ML disagree with each other, the appearance frequencies of number of elements in each material in four datasets are plotted in Figure 3c. One can see that in the exp. dataset used as the training set in this work, around 90% materials are binary compounds and 10% materials are ternary, while in the MP database there are about 40% materials that contain more than 3 elements. Since the training set doesn't cover materials space with more than 3 elements, the ML predictions for materials with more than 3 elements are extrapolations and in general less reliable than that for binary and ternary compounds. For the set of materials unstable by MP and stable by ML, the distribution of number of elements is similar to that of the exp. dataset where the majority of materials are binary or ternary, while in the set of materials stable by MP and unstable by ML, most materials have 4 or 5 elements. Here the lack of materials with more than 3 elements in the current $\Delta H_{\text{f}}^{\text{exp}}$ dataset suggests that the ML predictions for materials with more than 3 elements should be carefully checked if ML and MP disagree with each other, and it also suggests the urgency of building a comprehensive $\Delta H_{\text{f}}^{\text{exp}}$ dataset with sufficient entries of materials with more than 3 elements.

Table 3. Examples of materials that have novel physical properties and/or potential applications with $E_{\text{hull}}^{\text{MP}} > 0.16$ eV/atom and $E_{\text{hull}}^{\text{ML}} < 0.06$ eV/atom. The materials with experiment as one of the characterization methods are synthesized materials, and others are currently only hypothetical. $E_{\text{hull}}$ is in the unit of eV/atom.

| Materials | MP ID | $E_{\text{hull}}^{\text{MP}}$ | $E_{\text{hull}}^{\text{ML}}$ | Characterization method(s) | Comment/ novel physical property/ potential application |
|---|---|---|---|---|---|
| MnSnIr | mp-11480 | 0.5117 | 0 | Experiment | Largest difference between $E_{\text{hull}}^{\text{MP}}$ and $E_{\text{hull}}^{\text{ML}}$. |
| Ta$_3$Pb | mp-1187214 | 0.3386 | 0 | Experiment | Superconductor[84] |
| AgRh | mp-1183233 | 0.2633 | 0.0359 | Experiment | Electrocatalyst[70] |
| FeCoSn | mp-1025124 | 0.1836 | 0.0384 | Experiment | Tuning phase transitions for isostructural alloying[94] |
| SmCo$_4$Ag | mp-1219086 | 0.1797 | 0.0493 | Experiment | Positively correlated magnetization with temperature[78] |
| Li$_3$(FeS$_2$)$_2$ | mp-753818 | 0.1697 | 0.0180 | Experiment | Li-FeS$_2$ battery electrode[69] |
| PdRu | mp-1186459 | 0.2277 | 0.0032 | Experiment | Catalyst[71] |
| Ni$_3$Ag | mp-1100764 | 0.2332 | 0 | Experiment | Dual-frequency absorption[73] |
| Rb$_2$NaTaF$_6$ | mp-1114459 | 0.2038 | 0 | Experiment | Large anisotropic shift from both covalent and polarization spin transfer mechanisms[79] |
| Nb$_3$Tl | mp-569366 | 0.2083 | 0 | Experiment | Superconductor[83] |
| UPb$_3$ | mp-1184128 | 0.1621 | 0 | Experiment | Sharp metamagnetic transitions[80] |
| Cu$_3$N | mp-1933 | 0.1865 | 0.0464 | Experiment | Light recording media[74] |
| FeNi$_2$ | mp-1072076 | 0.1858 | 0.0292 | Experiment | Size-dependent catalytic activity[72] |
| HfCo$_7$ | mp-1105489 | 0.2098 | 0.0500 | Experiment | Rare-earth-free permanent magnets[81] |
| MnBi | mp-1185989 | 0.2078 | 0 | Experiment/DFT | Half-metallic ferromagnetism[77] |
| Be$_2$Si | mp-1009829 | 0.2352 | 0.0272 | Experiment/DFT | Hybrid nodal-line semimetal[76] |
| Mn$_2$Hg$_5$ | mp-30720 | 0.2362 | 0 | Experiment/DFT | π-based covalent magnetism[82] |
| Ta$_3$Bi | mp-1187199 | 0.3442 | 0 | DFT | Topological Dirac semimetal[91] |
| MnCrSb | mp-1221652 | 0.2564 | 0 | DFT | Half-metallicity[92] |
| LiB$_{11}$ | mp-1180507 | 0.2084 | 0.0234 | DFT | Pseudo-plasticity[90] |
| NiAg$_3$ | mp-976762 | 0.1850 | 0 | DFT | Acetylene adsorbent[89] |
| Li$_2$VN$_2$ | mp-1246112 | 0.1615 | 0.0279 | DFT | Li-ion battery electrode[87] |
| LiGdO$_3$ | mp-1185401 | 0.3476 | 0.0575 | Machine learning | Perovskite with high tolerance factor[88] |
| LiPmO$_3$ | mp-1185388 | 0.2815 | 0 | Machine learning | Perovskite with high tolerance factor[88] |

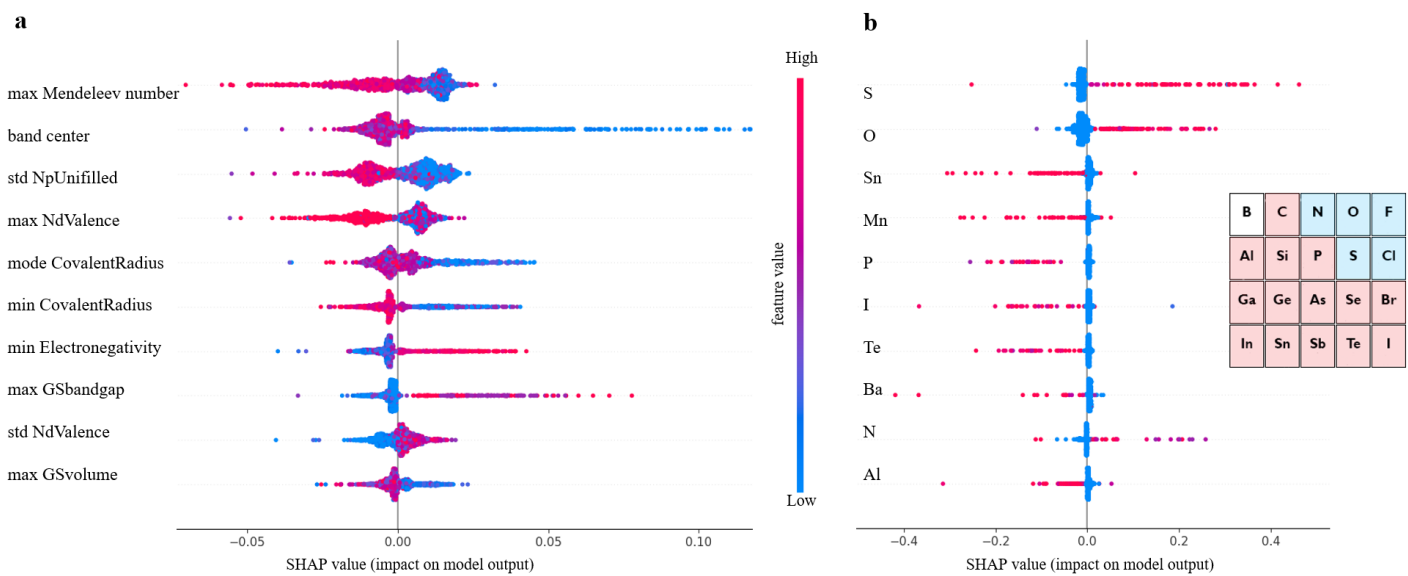

**Figure 4. Impact of each feature on model output. a** and **b** Distributions of the impacts (SHAP values[95]) of compositional features and elemental fractions on the model output ($\Delta H_f^{\text{diff}}$), respectively. The color represents the feature value (red high, blue low), and here only the top 10 features and elemental fractions with the highest sum of absolute SHAP values are shown. The inserted figure in **b** illustrates the trends of DFT to underestimate or overestimate $\Delta H_f$ of materials with certain non-metal elements. Here, the blue shaded elements are those for which DFT tends to underestimate $\Delta H_f$, the red shaded elements are those for which DFT tends to overestimate $\Delta H_f$, and Boron shows a mixed trend.

**Data-mining where $\Delta H_f^{\text{DFT}}$ fails by explaining the multi-fidelity model.** In addition to predicting more accurate $\Delta H_f$ and examining stability of materials, the random forest model trained on $\Delta H_f^{\text{diff}}$ ($\Delta H_f^{\text{exp}}$ – $\Delta H_f^{\text{DFT}}$) with human-engineered features can also serve as a data-mining approach to reveal where and how $\Delta H_f^{\text{DFT}}$ deviates from $\Delta H_f^{\text{exp}}$ (as above, "$\Delta H_f^{\text{DFT}}$" refers to the empirically corrected PBE $\Delta H_f$ by Jain *et al.*[46] in the Materials Project database), which provides clearer trends than machine learning models trained on $\Delta H_f^{\text{DFT}}$ only. Here, we analyze the relationship between human-understandable features and $\Delta H_f^{\text{diff}}$ by explaining the model, or for each material, calculating the impact of each feature on the model output (known as the SHAP value[95]). Previously, the error of $\Delta H_f^{DFT}$ is mostly discussed in the context of certain anions[3, 41, 46], cations[3] and transition metals[3, 41, 42]. In Figure 4a, the impacts of the top 10 compositional features from matminer[59] with the highest sum of absolute SHAP values are shown. One can see that, in addition to anion properties ("max GSbandgap", the detailed explanations of the descriptors are available in the matminer paper[59]) and cation properties ("max GSvolume", "max NdValence", "min CovalentRadius", "min Electronegativity"), mean field of elemental properties ("band center", "mode CovalentRadius") and standard deviation of elemental properties ("std NpUnifilled", "std NdValence") are also among the most impactful properties with respect to $\Delta H_f^{diff}$. For example, with smaller "band center" (geometric mean of electronegativity[59]), $\Delta H_f^{diff}$ tends to be larger and $\Delta H_f^{DFT}$ tends to be smaller than $\Delta H_f^{exp}$, which means that DFT tends to underestimate $\Delta H_f$ of systems with smaller mean electronegativity. A possible explanation for this trend is that, smaller geometric mean of electronegativity, the ability of atoms to bind the electrons near the atomic nuclei is weaker, and electrons tend to be delocalized. Since the GGA approximation tends to overestimate the electron delocalization[96], DFT (PBE) $\Delta H_f$ tends to be more negative for the systems with delocalized bonds (stronger bonding). Another example is, with larger standard deviation of number of *p* valence electrons, $\Delta H_f^{diff}$ tends to be smaller and $\Delta H_f^{DFT}$ tends to be larger than $\Delta H_f^{exp}$, suggesting that DFT tends to overestimate $\Delta H_f$ of systems with more dissimilar *p* valence electron configurations. This trend might be explained by the hypothesis that, with more different *p* electron configuration, in general the compound is more ionic, and because of the fact that GGA approximation tends to underestimate the electron localization[96], DFT (PBE) $\Delta H_f$ tends to be more positive for the systems with localized bonds (weaker bonding).

As for the impacts of certain cations and anions, or impacts of certain elements, we build a decision tree model that takes stoichiometry as input, and the SHAP values of fraction of each element are plotted in Figure 4b. One can see that, with higher atomic fraction of S, O and N, DFT tends to underestimate $\Delta H_f$, while for higher atomic fraction of Sn, Mn, P, I, Te, Ba, Al, DFT tends to overestimate $\Delta H_f$. There are more non-metal elements (6) in the top 10 most impactful elements than metals (2) and metalloids (2). Particularly, there is an interesting pattern of how DFT treats different non-metal elements: as shown in Figure 4b and Figure S2, for strong oxidizing non-metal elements in the upper-right corner of the periodic table, including F, O, N, S, Cl, DFT tends to underestimate $\Delta H_f$, while for those non-metal elements with weaker oxidizing ability, DFT tends to overestimate $\Delta H_f$. However, the degree of overestimation or underestimation doesn't simply correlate with the oxidizing ability. For example, as shown in Figure 4b and Figure S1, F has stronger oxidizing ability than O and S, but the degree of underestimation of $\Delta H_f^{DFT}$ for fluorides is less than that of oxides and sulfides. There are two possible sources of errors that would result in the observed trend: on the one hand, the underestimation or overestimation of $\Delta H_f$ of materials with certain elements might come from the element type-based empirical corrections[3, 46], and on the other hand, the intrinsic limit of the GGA and GGA + $U$ approximation might cause the different deviation patterns. For example, Seo et al.[97] proposed that the GGA + $U$ method used for transition metal oxides in the MP database[46] overestimates the degree of hybridization between the $d$ orbitals of transition metals and $p$ orbitals of oxygen, thus makes the calculated $\Delta H_f$ more negative.

The trend in Figure 4a also agrees with that in Figure 4b. For example, for "max GSbandgap" and "max GSvolume", they are calculated in the following procedure: first the ground state band gaps and ground state volumes of all the elements in the compound are listed, then the maximum values of band gaps and volumes are picked up. Therefore, "max GSbandgap" and "max GSvolume" actually relate to the existence of certain elements in the compound. Specifically, "max GSbandgap" describes the presence of specific anion in the compound while "max GSvolume" describes that of cation. Larger "max GSvolume", $\Delta H_f^{DFT}$ tends to be larger (more positive) than $\Delta H_f^{exp}$. An explanation for this trend is that with larger "max GSvolume", the cation element tends to have larger ground state volume (closer to the bottom-left of the periodic table with the maximum value at Cs). If the cation is closer to the bottom-left of the table, the compound in general will be more ionic. Therefore, $\Delta H_\text{f}^\text{DFT}$ tends to be more positive for the systems with more ionic bonds as mentioned above. On the other hand, larger "max GSbandgap", $\Delta H_\text{f}^\text{DFT}$ tends to be smaller (more negative) than $\Delta H_\text{f}^\text{exp}$. This phenomenon might be explained by the fact that, with larger "max GSbandgap", the anionic element is closer to the upper-right corner of the periodic table the maximum value at N, and according to Figure 4b the compound tends to have more negative $\Delta H_\text{f}^\text{DFT}$.

Note that in Wang *et al.*[47] all anionic corrections are negative, which is because their correction is applied to the original PBE results and PBE tends to overestimate the energy of diatomic gas molecules[98], while the trend shown here is based on the empirically corrected PBE energies from MP that already take the effect of overestimated energy of diatomic gas molecules into account.

## DISCUSSION AND CONCLUSION

In this work, we conduct a comprehensive machine learning study to learn and predict experimental formation enthalpy of materials. We use two different strategies to transfer information from larger DFT dataset to the smaller experimental dataset, transfer learning and multi-fidelity machine learning, and we use four machine learning architectures to realize the two strategies. We find that the random forest model trained on the difference between experimental and DFT formation enthalpy with DFT formation enthalpy as one of the input features can achieve the lowest error, which is half of that of DFT (empirically corrected PBE), and it also outperforms other more accurate but more computationally expansive density functionals, such as meta-GGA functionals. Beyond identifying the best model, we suggest that the deep neural network-based representation learning algorithms and transfer learning should not be the only machine learning architecture and information-transfer strategy considered. Other feature engineering methods such as human-engineered features, machine learning architectures beyond neural networks such as random forest and information-transfer strategy such as multi-fidelity machine learning should also be tested in machine learning applications for materials science.

As an application, we employ the found best random forest model to calibrate the formation enthalpy of all materials in the Materials Project database, which are then used to construct energy above hull and discover potential important materials that have underestimated stability in the MP database. Further, we use the machine learning model as a data-mining approach to identify patterns in the performance of DFT, for example in its tendency to underestimate the formation enthalpy of materials with elements in the upper-right corner of the periodic table.

Note that this work is based on the Materials Project database queried in March, 2021 (V2021.03.22). The methodology of this work can also be applied to updated Materials Project database (such as V2021.05.13) and other large DFT databases. It is expected that, with more accurate low fidelity data (DFT formation enthalpy), such as the recent dataset with 6,000 materials calculated by meta-GGA functionals[44], the method in this work can be used to provide more accurate calibration (exp. formation enthalpy).

One potential limitation of the multi-fidelity model used in this work is that it requires the availability of low-fidelity data for all the materials space of interest, as in this work DFT formation enthalpy is required for learning the difference of formation enthalpy from experiment and DFT, therefore it is required for predicting experimental formation enthalpy of materials in the Materials Project database. In cases where low-fidelity data is not available to all the materials, transfer learning might be more appropriate to transfer information between different datasets. Another scenario not considered in the current multi-fidelity machine learning scheme is that, for some properties there might be datasets with multiple levels of fidelity available. In such cases, in addition to incorporating different fidelity data into the input, the learning of differences might be conducted multiple times to enlarge the availability of high-fidelity data gradually.

## METHODS

**Data collection.** In this work, we construct the $\Delta H_f^{exp}$ dataset by combining two datasets from IIT[17] and SSUB[58], and we use the Materials Project[2] database (V2021.03.22) to construct the $\Delta H_f^{DFT}$ dataset. For the $\Delta H_f^{diff}$ dataset, since the $\Delta H_f^{DFT}$ values are provided for some materials in the IIT dataset, $\Delta H_f^{diff}$ values for those materials are obtained by subtracting the provided $\Delta H_f^{DFT}$ from the provided $\Delta H_f^{exp}$, and for materials from the SSUB dataset, since chemical formula is the only identifier, we take the lowest $\Delta H_f^{exp}$ for each formula, and for the $\Delta H_f^{DFT}$ of these materials, we assign the lowest $\Delta H_f^{DFT}$ to each formula. For overlaps between the IIT dataset and SSUB dataset, we take the $\Delta H_f^{exp}$ from the IIT database as the IIT database is a more recent one[17]. Note that the mean absolute difference of $\Delta H_f^{exp}$ between our dataset and the recent dataset from Wang *et al.* is only 0.007 eV/atom. Codes and a step-by-step instruction for constructing the dataset are provided in the **CODE AVAILABILITY** section.

**Machine learning models training procedure.** In this work, the dataset of the 1143 $\Delta H_f^{exp}$ is used for four purposes: 1) hyper-parameters tuning for each machine learning model, 2) model evaluation, and 3) production, or prediction of $\Delta H_f$ of all materials in the Materials Project database (MP). For purpose 1) and purpose 2), as mentioned in the **METHODS** section, we first randomly reserve 20% data as the test set for model selection (these 20% data are also excluded in the larger MP dataset for transfer learning). Then, to determine the best set of hyper-parameters for each model, with the remaining 80% data, we randomly reserve 20% of the remaining data (20%*80% = 16% of total data) as the validation set to evaluate each specific set of hyper-parameters, and use 80% of the remaining data (80%*80% = 64% of total data) to train the machine learning model with the given set of hyper-parameters. We screen hyper-parameters by grid search, and tables of search space of hyper-parameters are provided in the Supplementary Information. Finally, with the found best hyper-parameters for each model, we use the 80% of the data (training set + validation set in the hyper-parameter search step) to train machine learning models 10 times with different initialization, and evaluate model performance and uncertainty using the 20% data held out at the very beginning (test set). For purpose 3), production, for best prediction performance, all available 1143 data points are used to train the found best model with the found best hyper-parameter.

In this work, we use four different machine learning architectures to realize transfer learning and/or multi-fidelity machine learning, random forest (RF), multi-layer perceptron (MLP), Representation Learning from Stoichiometry (ROOST)[26] and Crystal Graph Convolutional Neural Network (CGCNN)[32]. For ROOST, we feed the compositions of materials as input, and it learns the representations of materials, and for CGCNN, we feed the 3D atomic structures of materials as input, and it also learns the representations. RF and MLP are realized by scikit-learn[99], and we use the descriptors from matminer[59] to feed RF and MLP as features of materials. Modules used to generate compositional features are ElementProperty, ElectronAffinity, BandCenter, CohesiveEnergy, Miedema, TMetalFraction, ValenceOrbital, YangSolidSolution, and modules used to generate structural features are GlobalSymmetryFeatures, StructuralComplexity, ChemicalOrdering, MaximumPackingEfficiency, MinimumRelativeDistances, StructuralHeterogeneity, AverageBondLength, AverageBondAngle, BondOrientationalParameter, CoordinationNumber, and DensityFeatures.

**Energy above hull.** In the Materials Project (MP), the energy above hull ($E_{hull}$) is defined as the energy of decomposition of a material into the set of most stable materials at this chemical composition[2]. The decomposition is tested against all potential chemical combinations that result in the material's composition. A positive $E_{hull}$ indicates that this material is unstable with respect to decomposition, and a zero $E_{hull}$ indicates that this compound is stable with respect to decomposition. In this work, the energy above hull is defined in the same way as MP. A graphical illustration of $E_{hull}$ is provided in Figure S2. In this work, the PhaseDiagram module in Pymatgen[100] is used to calculate the $E_{hull}$. The inputs required by the PhaseDiagram module are the compositions and formation enthalpies, and the corresponding output is the energies vs. compositions diagram, from which the decomposition energies and $E_{hull}$ can be calculated.

## DATA AVAILABILITY

All datasets, including the dataset of 1143 materials with experimental and MP formation enthalpies, the dataset of 122 materials used for comparison between different functionals with different corrections, and the dataset of 98,338 materials in the MP database with machine learning predicted formation enthalpies and energy above hull, are provided at:

https://doi.org/10.6084/m9.figshare.19100645.v1

## CODE AVAILABILITY

All scripts with their requirements and trained machine learning models are provided at:

https://doi.org/10.6084/m9.figshare.19100645.v1

# ACKNOWLEDGMENTS

This work was supported by Toyota Research Institute. Computational support was provided by the DOE Office of Science User Facility supported by the Office of Science of the U.S. Department of Energy under Contract No. DE-AC02-05CH11231, and the Extreme Science and Engineering Discovery Environment, supported by National Science Foundation grant number ACI-1053575.

# AUTHOR CONTRIBUTIONS

# COMPETING INTERESTS

12. Zhu H, *et al.* Computational and experimental investigation of TmAgTe2 and XYZ2 compounds, a new group of thermoelectric materials identified by first-principles high-throughput screening. *J Mater Chem C* **3**, 10554-10565 (2015).

13. Dunstan MT, *et al.* Large scale computational screening and experimental discovery of novel materials for high temperature $CO_2$ capture. *Energy Environ Sci* **9**, 1346-1360 (2016).

14. Li S, *et al.* Data-Driven Discovery of Full-Visible-Spectrum Phosphor. *Chem Mater* **31**, 6286-6294 (2019).

15. Cooley JA, Horton MK, Levin EE, Lapidus SH, Persson KA, Seshadri R. From Waste-Heat Recovery to Refrigeration: Compositional Tuning of Magnetocaloric $Mn_{1+x}Sb$. *Chem Mater* **32**, 1243-1249 (2020).

16. Perdew JP, Burke, K., & Ernzerhof, M. Generalized Gradient Approximation Made Simple. *Phys Rev Lett* **77**, 3865 (1996).

17. Kim G, Meschel SV, Nash P, Chen W. Experimental formation enthalpies for intermetallic phases and other inorganic compounds. *Sci Data* **4**, 170162 (2017).

18. Bartel CJ, Trewartha A, Wang Q, Dunn A, Jain A, Ceder G. A critical examination of compound stability predictions from machine-learned formation energies. *npj Comput Mater* **6**,  (2020).

19. Aykol M, Dwaraknath, S. S., Sun, W., & Persson, K. A. Thermodynamic limit for synthesis of metastable inorganic materials. *Sci Adv* **4**, eaaq0148 (2018).

20. Sun W, Dacek, S.T., Ong, S.P., Hautier, G., Jain, A., Richards, W.D., Gamst, A.C., Persson, K.A. and Ceder, G. The thermodynamic scale of inorganic crystalline metastability. *Sci Adv* **2**, e1600225 (2016).
**32 / 41**